# THE SIZE IMPACT ON EVOLUTIONARY DYNAMICS OF AGENTS-CLUSTERS

*Vitalie Eremeev*[1,2], *Ion Rasca*[1] *and Florentin Paladi*[1]

[1]**Department of Theoretical Physics, State University of Moldova**
A.Mateevici str.60, Chisinau MD-2009, Moldova

[2]**Department of Engineering and Computer Sciences**
Free International University of Moldova
Vlaicu Parcalab str.52, Chisinau MD-2012, Moldova

E-mail: fpaladi@usm.md

**Abstract**

A cluster theory based mathematical model was developed and used to simulate the dynamics of a system composed of a large number of interacting agents-clusters with different size. The case of a system formed by a constant total number of economic units (agents) in metastable (partial) equilibrium was considered, and the size effect on the formation of the groups (clusters) of agents was particularly elucidated. We prove that the fragmentation and coagulation rates of groups of agents definitely depend on the size of the group.

# THE SIZE IMPACT ON EVOLUTIONARY DYNAMICS OF AGENTS-CLUSTERS

*Vitalie Eremeev[1,2], Ion Rasca[1] and Florentin Paladi[1*]*

[1]*Department of Theoretical Physics, State University of Moldova,*
*A.Mateevici str.60, Chisinau MD-2009, Moldova*

[2]*Department of Engineering and Computer Sciences,*
*Free International University of Moldova*
*Vlaicu Parcalab str.52, Chisinau MD-2012, Moldova*

A cluster theory based mathematical model was developed and used to simulate the dynamics of a system composed of a large number of interacting agents-clusters with different size. The case of a system formed by a constant total number of economic units (agents) in metastable (partial) equilibrium was considered, and the size effect on the formation of the groups (clusters) of agents was particularly elucidated. We prove that the fragmentation and coagulation rates of groups of agents definitely depend on the size of the group.

**1. Introduction**

The mathematical modeling in economics is a very important tool for investigation of economic and financial processes. In particular, many papers were recently published in the field of application of the methods of statistical physics, thermodynamics, synergetics etc. in economics [1-5], and they are on the increase. Each economic sector formed by many agents that interact reciprocally has some similar features with physical systems which are composed by many different components or so-called *cluster*-structures. In general, a cluster represents an aggregate formed by a number of similar elements that are thought to be unitary and indivisible.

Meanwhile, the notion of economic or financial cluster, which assumes, for instance, the division of market in similar elements, can be used in mathematical modeling. The future segmentation of the market can be done regarding these components, and the examination of each segment as an aggregate of clusters of a certain size in the units of reference used initially in the process of separation can be justified by the fact that the economic systems have a complex evolving structure, e.g. the firms competing in an economy are composed by divisions. In the case of a perfect competition, in order to avoid the appearance of non-aggregative processes, we shall consider a closed system of economic or financial structures, without the input/output flows, i.e. a system formed by a constant total number of economic or financial agents in partial equilibrium which supposes the existence of some possibilities for a new optimum redistribution of money and goods on the market. This situation is analogous to the realization of a local minimum of the Gibbs free energy in the cluster theory [1, 6]. Thus, a stable economic development assumes an evolution of the system to such partition of money and goods which corresponds to the deeper thermodynamic minimum of energy, that is the truly stable equilibrium state of the system.

The notion of cluster is used in the next paragraphs with the meaning of a complex structure formed by a number of similar economic or financial units which interact on the market within a closed system of structures (agents) with different size. Analytical and numerical study of the influence of the size of the clusters on their formation and evolution in conditions of partial equilibrium and perfect competition is the aim of this paper.

---

[*] - contact author. E-mail: fpaladi@usm.md

## 2. Mathematical model

In the framework of the cluster approach the following two basic assumptions, which allow a mathematical formalism to be developed for a detailed description of the evolution, are indispensable [6]:

(i) There exist clusters in the initial state which consist of different number $n$ of elements ($n=1, 2, \ldots$),
(ii) Transformations of $n$-sized clusters into $m$-sized ones at time $t$ occur with certain, in general, time-dependent frequencies $f_{nm}(t)(s^{-1})$ ($n, m=1, 2, \ldots$).

The evolution of the process is sought to be described by function $Z_n(t)$, which represents the solution of the kinetic master equation and characterizes the time-dependence of the concentration of clusters of size $n$. Figure 1 shows schematically how $n$-sized cluster can increase or decrease its size. In particular, the arrow beginning from size $n$ and ending at size $m$ on the size axis symbolizes the quantity $f_{nm}(t)Z_n(t)$, which gives the number of $n \to m$ transitions undergone by the $n$-sized clusters per unit time, divided by the total number of interacting agents-clusters with different size $N$. Then, the concentration of clusters of size $n$ will be diminished per unit time by the quantity:

$$\sum_{m=1}^{M} f_{nm}(t)Z_n(t), \qquad (1)$$

where $M$ is the total number of economical units (agents) existing in the system. Conversely, the arrows ending at size $n$ illustrate the role of the reverse, i.e. the $m \to n$, transitions: owing to them $Z_n(t)$ will increase per unit time by the quantity:

$$\sum_{m=1}^{M} f_{mn}(t)Z_m(t). \qquad (2)$$

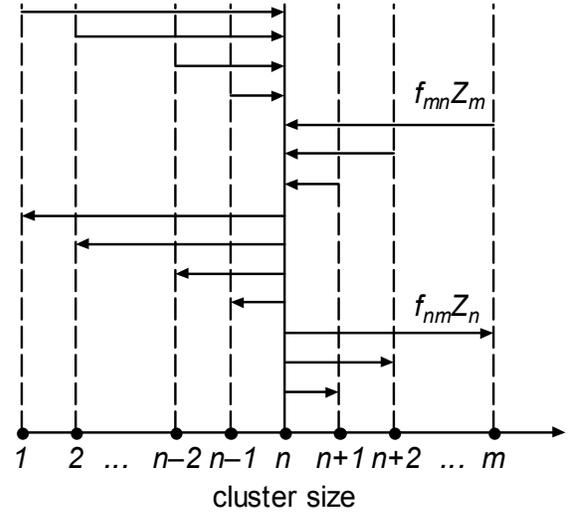

Fig.1. Schematic presentation of the possible changes in the size of a cluster of $n$ units.

On the other hand, the change of $Z_n(t)$ per unit time is expressed mathematically by the derivative $dZ_n(t)/dt$. The balance between the above quantities thus leads to the sought master equation for a closed system:

$$\frac{dZ_n(t)}{dt} = \sum_{m=1}^{M}\left[f_{mn}(t)Z_m(t) - f_{nm}(t)Z_n(t)\right]. \qquad (3)$$

Equation (3) is a set of ordinary differential equations of first order. In general, these equations are non-linear because of the dependence of the transition frequencies on the unknown cluster concentration $Z_n(t)$. Clearly, $Z_n(t)$ and $M$ are connected by the relation $\sum_{n=1}^{M} nZ_n(t) = M/N$, and the initial cluster size distribution $Z_n(0)$ is considered to be *a priori* known.

We can now study easily the processes of type $[Z_1]+[Z_n] \leftrightarrow [Z_{n+1}]$, occurring especially at the early stages, when it is unlikely for the clusters of $n=2, 3, \ldots$ units to interact, because their concentration is still rather low. This is illustrated in Figure 2 in which the arrows symbolize the number of forward ($n{\rightarrow}n{+}1$) and backward ($n{\rightarrow}n{-}1$) transitions. Denoting $f_n=f_{n,n+1}(t)$, $f_{n-1}=f_{n-1,n}(t)$, $g_n=f_{n,n-1}(t)$, $g_{n+1}=f_{n+1,n}(t)$, where $f_{nm}(t)=0$ for $|n-m|>1$ and $f_{nm}(t)\neq 0$ only for $|n-m|=1$, the clusters will change size by nearest-size transitions and the equation (3) becomes:

$$\frac{dZ_n(t)}{dt} = f_{n-1}Z_{n-1}(t) - g_n Z_n(t) - f_n Z_n(t) + g_{n+1}Z_{n+1}(t) . \qquad (4)$$

By definition, $f_0=0$, $g_1=0$ and $Z_{M+1}=0$. The initial condition for a closed system ($M=const$) in partial equilibrium, when at the initial moment $t=0$ there are only economical units in the system whose concentration $Z_1$ equals their equilibrium concentration $C_1$, is $Z_1(0)=C_1$, and $Z_n(0)=0$ ($n=2, 3, \ldots, M$). The boundary conditions for the unknown cluster size distribution, in conformity with the assumption that the actual concentration of units in a system in partial equilibrium is equal to the equilibrium one, are $Z_1(t)=C_1$ and $Z_M(t)=0$. We shall require $Z_1(t)$ to remain constant with time, and this assumption is based on the fact that $Z_1$ and $C_1$ are thought to be large numbers. Thus, equation (4) becomes a set of $M$-2 ordinary linear differential equations of first order in the $M$-2 unknowns $Z_2(t), Z_3(t), \ldots, Z_{M-1}(t)$.

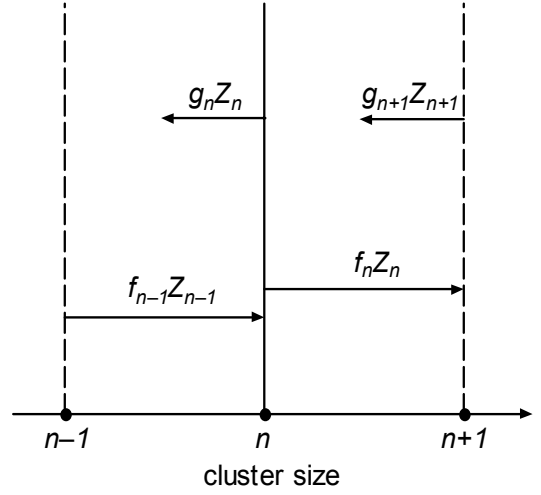

Fig.2. Schematic presentation of the possible changes in the size of a cluster of $n$ units by nearest-size transitions.

The first step in solving equation (4) is to homogenize it by presenting $Z_n(t)$ in form of the sum of stationary cluster size distribution, $X_n$, and unknown deviation of $Z_n(t)$ from $X_n$, $y_n(t)$:

$$Z_n(t) = X_n + y_n(t) , \qquad (5)$$

where $y_n(t)$ satisfies the initial condition $y_n(0)=-X_n$, and $dy_n(t)/dt = f_{n-1}y_{n-1}(t) - (f_n+g_n)y_n(t) + g_{n+1}y_{n+1}(t)$ ($n=2, 3, \ldots, M$-1) which is a set of $M$-2 already homogeneous ordinary linear differential equations of first order with time-independent coefficients. For any fixed $n=2, 3, \ldots, M$-1 we shall have $M$-2 linearly independent particular solutions $y_{ni}(t)$ of the form:

$$y_{ni}(t) = a_{ni}\exp(-\lambda_i t) \quad (i=2, 3, \ldots, M-1), \qquad (6)$$

where $a_{ni}$ are constants (to ensure the linear independence of $y_{ni}(t)$ we shall require that not all $a_{ni}$ equal zero simultaneously), and $\lambda_i>0$ is the $i$th eigenvalue, i.e. the $i$th root of the characteristic equation

$$\begin{vmatrix} f_2+g_2-\lambda & -g_3 & 0 & \cdots & 0 & 0 & 0 \\ -f_2 & f_3+g_3-\lambda & -g_4 & \cdots & 0 & 0 & 0 \\ 0 & -f_3 & f_4+g_4-\lambda & \cdots & 0 & 0 & 0 \\ \vdots & \vdots & \vdots & \vdots & \vdots & \vdots & \vdots \\ 0 & 0 & 0 & \cdots & f_{M-3}+g_{M-3}-\lambda & -g_{M-2} & 0 \\ 0 & 0 & 0 & \cdots & -f_{M-3} & f_{M-2}+g_{M-2}-\lambda & -g_{M-1} \\ 0 & 0 & 0 & \cdots & 0 & -f_{M-2} & f_{M-1}+g_{M-1}-\lambda \end{vmatrix} = 0$$

The above determinant represents a polynomial of degree $M$-2 which has $M$-2 simple roots $\lambda_2, \lambda_3, \ldots, \lambda_{M-1}$. The next step is, therefore, to find these roots, and then to determine, for each $i=2, 3, \ldots, M$-1, the constants $a_{ni}$ with the help of the recursion formulae:

$$(f_2+g_2-\lambda_i)a_{2i} - g_3 a_{3i} = 0,$$
$$-f_{n-1}a_{n-1,i} + (f_n+g_n-\lambda_i)a_{ni} - g_{n+1}a_{n+1,i} = 0, \quad (n=3,4,\ldots,M-2) \qquad (7)$$

$$-f_{M-2}a_{M-2,i}+(f_{M-1}+g_{M-1}-\lambda_i)a_{M-1,i}=0,$$

where it is convenient to set, without loss of generality, $a_{M-1,i}=1$ for each $i=2, 3, \ldots, M-1$.

We can now use the $M-2$ linearly independent solutions $y_{ni}(t)$ from (6) in order to represent the general solution $y_n(t)$ as a linear combination of them:

$$y_n(t) = \sum_{i=2}^{M-1} c_i a_{ni} \exp(-\lambda_i t) \quad (n=2, 3, \ldots, M-1). \tag{8}$$

The last step is to find the $M-2$ unknown constants $c_i$ which are the solution of the linear algebraic set of $M-2$ equations resulting from using the initial condition for $y_n$ in (8):

$$\sum_{i=2}^{M-1} c_i a_{ni} = -X_n \quad (n=2, 3, \ldots, M-1). \tag{9}$$

In accordance with the Cramer rule, $c_i$ is given by $c_i=d_i/d'$ ($i=2, 3, \ldots, M-1$), where $d_i$ and $d'$ are the following determinants of order $M-2$:

$$d_i = \begin{vmatrix} a_{22} & a_{23} & \cdots & a_{2,i-1} & -X_2 & \cdots & a_{2,M-1} \\ a_{32} & a_{33} & \cdots & a_{3,i-1} & -X_3 & \cdots & a_{3,M-1} \\ \vdots & \vdots & \vdots & \vdots & \vdots & \vdots & \vdots \\ a_{M-1,2} & a_{M-1,3} & \cdots & a_{M-1,i-1} & -X_{M-1} & \cdots & a_{M-1,M-1} \end{vmatrix}, \tag{10}$$

$$d' = \begin{vmatrix} a_{22} & a_{23} & \cdots & a_{2,i-1} & a_{2i} & \cdots & a_{2,M-1} \\ a_{32} & a_{33} & \cdots & a_{3,i-1} & a_{3i} & \cdots & a_{3,M-1} \\ \vdots & \vdots & \vdots & \vdots & \vdots & \vdots & \vdots \\ a_{M-1,2} & a_{M-1,3} & \cdots & a_{M-1,i-1} & a_{M-1,i} & \cdots & a_{M-1,M-1} \end{vmatrix}. \tag{11}$$

Using (9) and inserting $y_n(t)$ from (8) into (5), we find the solution of the problem:

$$Z_n(t) = X_n + \sum_{i=2}^{M-1} (d_i/d') a_{ni} \exp(-\lambda_i t) \tag{12}$$

or, equivalently,

$$Z_n(t) = X_n \left\{ 1 - \left[ \sum_{i=2}^{M-1} d_i a_{ni} \right]^{-1} \sum_{i=2}^{M-1} d_i a_{ni} \exp(-\lambda_i t) \right\}, \tag{13}$$

where $X_n = \dfrac{f_1 f_2 \cdots f_{n-1}}{g_2 g_3 \cdots g_n} \left[ 1 + \sum_{m=2}^{M-1} \dfrac{g_2 g_3 \cdots g_m}{f_2 f_3 \cdots f_m} \right]^{-1} \sum_{m=n}^{M-1} \dfrac{g_2 g_3 \cdots g_m}{f_2 f_3 \cdots f_m}$ ($n=2, 3, \ldots, M-1$) (14)

is the stationary cluster size distribution divided by its equilibrium value $X_1$.

Equations (12) or (13) represent the time-dependent cluster size distribution which has an origin at no preexisting clusters in the system. We note that, as it should be, $Z_n(t) \to X_n$ for $t \to \infty$. Meanwhile, the exact expressions for the roots $\lambda_i$ can be found analytically only for $M-2 \leq 4$, i.e. for $M \leq 6$. For $M>6$ we must resort to numerical methods for solving the problem.

## 3. Results and discussions

The relative increase with time of the quantity of $n$-sized agents-clusters was numerically simulated and the results for $M=7$ ($n=2, 3, \ldots, 6$) and $M=12$ ($n=2, 3, \ldots, 11$) are shown in Figures 3 and 4, respectively. We can conclude that the functions that approximate analytically the curvatures presented in these figures are different. In the time limit $t \rightarrow 0$ these functions change accordingly to the power law $Z_n(t) \sim t^{n-1}$, regardless of the value of $M$, and the relative time dependence of $Z_n(t)/X_n$ does not depend on the frequency $f_1$. Meanwhile, the evolution of any $n$-sized agent-cluster depends strongly on the total number of economical units (agents) existing in the system.

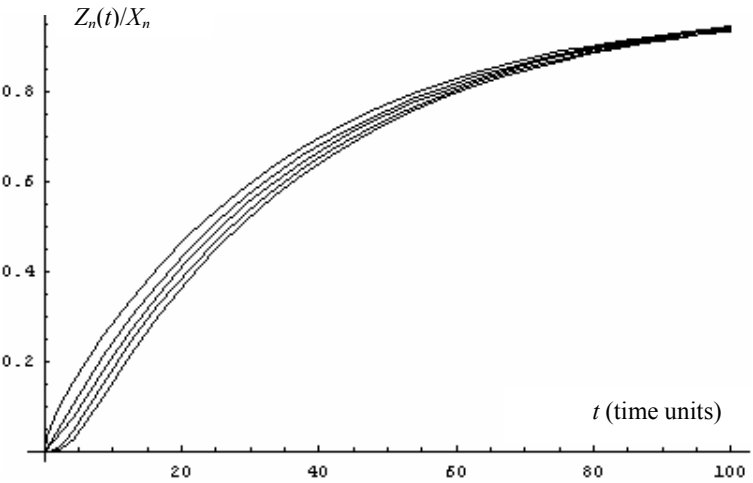

Fig.3. Relative time dependence of the quantity of $n$-sized agents-clusters ($M$=7); $n$=2, 3, 4, 5 and 6, from top to bottom.

Assigning the agents-clusters with sizes $n$=2, 3, and even 4 to the small business sector of the market, it is worth mentioning the existence of a linear or near linear evolution of the functions for the short-time periods. These structures reach a stable equilibrium faster than any larger agent-cluster ($n$>4). Therefore, even for a poorly advanced economy, the small economic agents can quickly increase in the number and, finally, reach the value which corresponds to the equilibrium one.

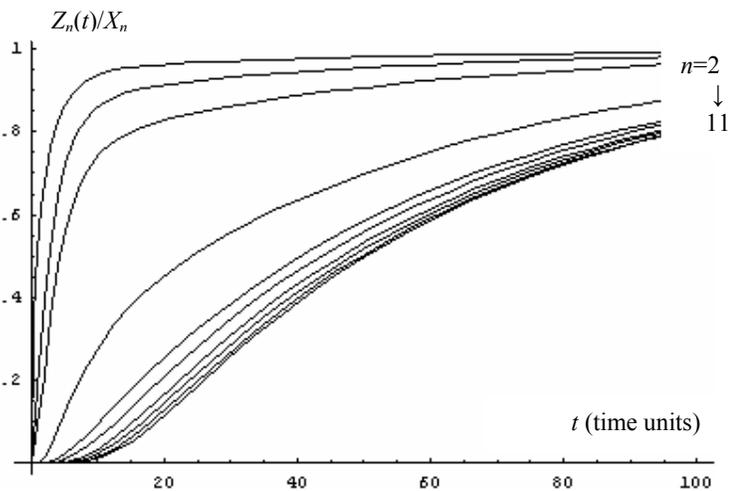

Fig.4. Relative time dependence of the quantity of $n$-sized agents-clusters for $M$=12; $n$=2, 3, …, 11, from top to bottom.

The further thriving of the small business requires new modifications on the entire market by increasing, for example, the investments or the quantity of larger agents in order to create a new metastable (partial) equilibrium on the market as a prerequisite for the continuation of the development of small agents.

Our results are in good agreement with the current reports on the economy of the Republic of Moldova [7].

## 4. Conclusion

A cluster theory based mathematical model was developed and applied qualitatively to the study of the size dependence of the development of relatively small economic agents-clusters, and for short-time periods we proved that the fragmentation and coagulation rates of groups of agents depend on their size.

The model could be developed further for the behavior of large groups of agents in a more complex environment which is closer to an actual economic situation, as well as for the presence of the input and output flows of money and goods on the market.

**References**

1. Saslow W.M., An economic analogy to thermodynamics, Am. J. Phys., 1999, V.67, No.12, P.1239-1247.
2. Amaral L.A.N., Buldyrev S.V., Havlin S., Salinger M.A. and Stanley H.E., Power law scaling for a system of interacting units with complex internal structure, Phys. Rev. Lett., 1998, V.80, No.7, P.1385-1388.
3. Zheng D., Rodgers G.J. and Hui P.M., A model for the size distribution of customer groups and businesses, Physica A, 2002, V.310, P.480-486.
4. Mantegna R.N., Palagyi Z. and Stanley H.E., Applications of statistical mechanics to finance, Physica A, 1999, V.274, P.216-221.
5. Chakraborti A. and Chakrabarti B.K., Statistical mechanics of money: How saving propensity affects its distribution, arXiv:cond-mat/0004256, V.2, 2000.
6. Kashchiev D., Nucleation: basic theory with applications, Butterworth-Heinemann, 2000, 529 P.
7. Paladi F., Rasca I. and Eremeev V., The size effect on economic development: a new cluster theory based model, Economica (Chisinau edition, in Romanian), 2003, No.1 (41), P.88-93.